\documentclass[preprint,showpacs,preprintnumbers,amsmath,amssymb]{revtex4}
\usepackage{graphics}
\usepackage{dcolumn}
\usepackage{bm}

\begin{document}

\title{Experimental evidence for new symmetry axis of electromagnetic beams}

\author{Chun-Fang Li\footnote{Email address: cfli@shu.edu.cn}}

\affiliation{Department of Physics, Shanghai University, 99 Shangda Road, Baoshan District, 200444
Shanghai, P. R. China}

\date{\today}

\begin{abstract}

The new symmetry axis of a well-behaved electromagnetic beam advanced in paper Physical Review A
78, 063831 (2008) is not purely a mathematical concept. The experimental result reported by Hosten
and Kwiat in paper Science 319, 787 (2008) is shown to demonstrate the existence of this symmetry
axis that is neither perpendicular nor parallel to the propagation axis.

\end{abstract}

\pacs{41.20.Jb, 02.10.Yn, 41.90.+e, 42.25.Ja}
\maketitle


\textit{Introduction.}---Finite electromagnetic beams have now played very important roles in
diverse areas of applications such as optical trapping and manipulation \cite{Ashkin}, optical
rotating \cite{Paterson}, optical guiding \cite{Xu-KJK}, optical data storage, and dark-field
imaging \cite{Biss-YB}. But an incredible fact is that the theoretical description of a
well-behaved finite beam has not been satisfactory \cite{Lax-LM, Davis, Pattanayak-A, Davis-P,
Jordan-H, Youngworth-B, Onoda-MN, Bliokh-B, Li1} since the advent of masers and lasers
\cite{Green-W, Kogelnik, Kogelnik-L}. It is well known that the vectorial property of a plane wave
is described by the polarization state. A useful concept is the Jones vector that consists of the
two mutually orthogonal transverse components \cite{Jones}. But for a finite beam, the
polarization state is not a global property \cite{Pattanayak-A}. Instead it is local and changes
on propagation. As a matter of fact, due to the longitudinal component \cite{Lax-LM, Li1}, the
polarization state is not sufficient to describe the vectorial property of a finite beam.
Recently, I advanced a new symmetry axis \cite{Li2} represented by a unit vector $\mathbf{I}$ and
found that it is a global property and, together with the Jones vector of the angular spectrum,
can describe the vectorial property of a beam. The purpose of this paper is to show that this
symmetry axis is not purely a mathematical concept. In fact, a perpendicular $\mathbf{I}$ to the
propagation axis corresponds to the uniformly polarized beam in the paraxial approximation
\cite{Davis, Pattanayak-A}, and a parallel $\mathbf{I}$ corresponds to the cylindrical vector beam
\cite{Davis-P, Youngworth-B, Li1}. The conversion from uniformly polarized beams to cylindrical
vector beams has been experimentally realized \cite{Youngworth-B, Ren-LW}. In this paper, I will
show that a symmetry axis $\mathbf{I}$ that is neither perpendicular nor parallel to the
propagation axis has been observed by Hosten and Kwiat \cite{Hosten-K} in a recent experiment on
the Imbert-Fedorov effect \cite{Fedorov, Imbert}.

\textit{Transverse displacement.}---In order to see the connection of $\mathbf{I}$ with the
Imbert-Fedorov effect, let me first introduce the transverse effect \cite{Li2}, the displacement
of the beam's barycenter from the characteristic plane formed by $\mathbf{I}$ and the propagation
axis, in the linear approximation. The electric vector $\mathbf{F}(\mathbf{x})$ of a monochromatic
beam traveling in positive $z$-axis is expressed as an integral over the angular spectrum,
\begin{equation} \label{integral}
\mathbf{F}(\mathbf{x})= \frac{1}{2\pi} \int_{k^2_x+k^2_y \leq k^2} \mathbf{f} (k_x,k_y)
\exp(i\mathbf{k} \cdot \mathbf{x}) dk_x dk_y,
\end{equation}
where $k_z= (k^2- k^2_x- k^2_y)^{1/2}$, and the electric vector $\mathbf{f}$ of the angular
spectrum is factorized as the following form,
\begin{equation} \label{factorization}
\mathbf{f}(k_x,k_y)=m \tilde{\alpha} f(k_x,k_y).
\end{equation}
The mapping matrix $m$ is so defined in terms of $\mathbf{I}$ and the wave vector $\mathbf{k}$
that it is normalized as $m^T m=1$, where the superscript $T$ stands for the transpose. As in Ref.
\cite{Li2}, $\mathbf{I}$ is set to lie in the plane $zox$ and to make an angle $\Theta$ with the
propagation axis, $ \mathbf{I}(\Theta)= \mathbf{e}_z \cos \Theta+ \mathbf{e}_x \sin \Theta$, where
$|\Theta| \leq \frac{\pi}{2}$ is assumed. The normalized Jones vector
\begin{equation} \label{Jones vector}
\tilde{\alpha}= \left( \begin{array}{c} \alpha_p\\\alpha_s \end{array} \right)
\end{equation}
is assumed to be the same to all the elements of the angular spectrum. The amplitude scalar
$f(k_x, k_y)$ of the angular spectrum is assumed here to include a phase factor,
\begin{equation} \label{scalar function}
f(k_{\rho}, \varphi)= f_0 (k_{\rho}) \exp(i l \varphi),
\end{equation}
in circular cylindrical system, where $f_0 (k_{\rho})$ is square integrable and is sharply peaked
at $k_{\rho}=0$, and $l$ is an integer. The transverse displacement $y_b$ of the beam's barycenter
is defined as the expectation of the $y$ coordinate,
\begin{equation} \label{definition of barycenter}
y_b= \langle y \rangle= \frac{\int \int \mathbf{F}^{\dag} y \mathbf{F} dx dy}{\int \int
\mathbf{F}^{\dag} \mathbf{F} dx dy},
\end{equation}
where the superscript $\dag$ stands for the conjugate transpose. Substituting Eqs.
(\ref{integral}), (\ref{factorization}), and (\ref{scalar function}) into Eq. (\ref{definition of
barycenter}) and noticing that $f_0$ is an even function of $k_y$, one get
\begin{equation} \label{TD integral}
y_b= i \frac{\int \int \tilde{\alpha}^{\dag} m^T \frac{\partial m} {\partial k_y} \tilde{\alpha}
|f_0|^2 dk_x dk_y}{\int \int |f_0|^2 dk_x dk_y}.
\end{equation}
It is noted that this displacement is independent of the phase factor $\exp(i l \varphi)$ in the
scalar function (\ref{scalar function}). When $|\Theta| \gg \delta \theta$, where $\delta \theta$
is half the divergence angle, $m$ is linearly approximated as \cite{Li2}
\begin{equation} \label{linear MM}
m= \mathrm{sgn}(\Theta)
  \left(
        \begin{array}{cc}
                 1                           &  \frac{k_y}{k} \cot \Theta\\
               -\frac{k_y}{k} \cot \Theta    &   1 \\
               -\frac{k_x}{k}                & -\frac{k_y}{k}
        \end{array}
  \right).
\end{equation}
Substituting Eqs. (\ref{Jones vector}) and (\ref{linear MM}) into Eq. (\ref{TD integral}), we have
\cite{Li2}
\begin{equation} \label{displacement}
y_b= -\frac{\sigma}{k} \cot \Theta,
\end{equation}
where $\sigma= \tilde{\alpha}^{\dag} \hat{\sigma} \tilde{\alpha}$ is the polarization ellipticity
of the angular spectrum, and $\hat{\sigma}= \left( \begin{array}{cc} 0 & -i \\ i & 0 \end{array}
\right)$ is the Pauli matrix.

\begin{figure}[ht]
\includegraphics{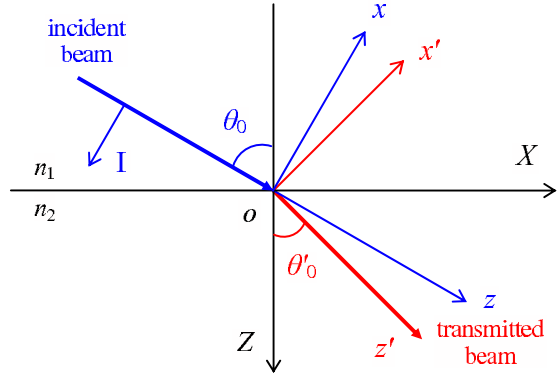}
\caption{(Color online) Reference frames $xyz$, $x'y'z'$, and $XYZ$ associated, respectively, with
the incident beam, the refracted beam, and the laboratory for the refraction at an interface
between two different dielectric media, $n_1$ and $n_2$.} \label{coordinates}
\end{figure}

\textit{Imbert-Fedorov effect: change of $\mathbf{I}$ by refraction.}---Next I will show that the
observed Imbert-Fedorov effect in Ref. \cite{Hosten-K} demonstrates the change of $\mathbf{I}$ by
the refraction. Consider the refraction at an interface between two different dielectric media,
$n_1=1$ and $n_2=1.515$, as is depicted in Fig. \ref{coordinates}, where the laboratory reference
frame is denoted by $XYZ$, the reference frame associated with the incident beam is denoted by
$xyz$, the reference frame associated with the transmitted beam is denoted by $x'y'z'$, $\theta_0$
and $\theta'_0$ are the incidence and refraction angles of the propagation axis, respectively. The
symmetry axis $\mathbf{I}$ of the incident beam in Ref. \cite{Hosten-K} is perpendicular to the
propagation axis. Let us start with an incident beam the perpendicular $\mathbf{I}$ of which lies
in the incidence plane.

\textit{1. Description of the incident beam.} In frame $xyz$, the wave vector of the angular
spectrum in the linear approximation is given by $ \mathbf{k}= (\begin{array}{ccc} k_x & k_y & k
\end{array})^T$, where the wave number in medium $n_1$ is $k= n_1 k_0$, $k_0= 2 \pi/ \lambda_0$,
and $\lambda_0$ is the vacuum wavelength. With $\Theta= -\frac{\pi}{2}$, the linearly approximated
mapping matrix (\ref{linear MM}) becomes
\begin{equation} \label{incidence mm}
m= -
  \left(
        \begin{array}{cc}
                 1            &  0\\
                 0            &  1 \\
               -\frac{k_x}{k} & -\frac{k_y}{k}
        \end{array}
  \right).
\end{equation}
In frame $XYZ$, the wave vector is transformed into
\begin{equation}\label{incident wave vector in XYZ}
\mathbf{k}_{XYZ}=D_y (-\theta_0) \mathbf{k}
  =\left(\begin{array}{c} k\sin\theta_0+k_x \cos\theta_0 \\
                     k_y                                \\
                     k\cos\theta_0-k_x \sin\theta_0
         \end{array}
   \right),
\end{equation}
where
$$
D_y (\vartheta)=\left( \begin{array}{ccc}
                                 \cos\vartheta & 0 & -\sin\vartheta\\
                                 0          & 1 &  0         \\
                                 \sin\vartheta & 0 &  \cos\vartheta\\
                    \end{array}
             \right)
$$
is the rotation matrix of the reference frame around the $y$ axis by an angle $\vartheta$. The
mapping matrix is correspondingly transformed into $m_{XYZ}= D_y (-\theta_0) m$ and is given by
\begin{equation} \label{incidence MM in XYZ}
m_{XYZ}= -
  \left(
        \begin{array}{cc}
                \cos\theta_0-\frac{k_x}{k}\sin\theta_0 & -\frac{k_y}{k}\sin\theta_0\\
                0                                            &  1                            \\
               -\sin\theta_0-\frac{k_x}{k}\cos\theta_0 & -\frac{k_y}{k}\cos\theta_0
        \end{array}
  \right).
\end{equation}

\textit{2. Operations of the refraction to the incident angular spectrum.} Refraction includes two
kinds of operation to the incident angular spectrum. One is to alter the wave vector and the
mapping matrix by rotation, and the other is to change the Jones vector. So the electric vector of
the refracted angular spectrum in the frame $XYZ$ can be assumed to be
\begin{equation}\label{field vector in XYZ}
\mathbf{f}'_{XYZ}= m'_{XYZ} \tilde{\alpha}' f',
\end{equation}
where $\tilde{\alpha}'$ is also normalized. Let us look at the former operation in frame $XYZ$.
For an incident plane wave of wave vector (\ref{incident wave vector in XYZ}), the incidence plane
is formed by the wave vector (\ref{incident wave vector in XYZ}) and the $Z$ axis. So the
incidence angle $\theta$ of this plane wave is determined by
$$
\cos\theta=\frac{\mathbf{k}_{XYZ} \cdot \mathbf{e}_Z}{|\mathbf{k}_{XYZ}|}
  =\cos\theta_0-\frac{k_x}{k} \sin\theta_0.
$$
Denoting by $\theta'$ the refraction angle, the Snell law gives
$$
\sin \theta'=\frac{n_1}{n_2} \sin\theta= \sin\theta'_0+ \frac{n_1}{n_2} \frac{k_x}{k}
\cos\theta_0.
$$
The unit vector $\mathbf{n}$ perpendicular to this incidence plane is given by
$$
\mathbf{n}=\frac{\mathbf{e}_Z \times \mathbf{k}_{XYZ}}{|\mathbf{e}_Z \times \mathbf{k}_{XYZ}|}
\approx \mathbf{e}_y-\frac{k_y}{k\sin\theta_0} \mathbf{e}_x
$$
in the linear approximation. Denoting $\Delta\theta= \theta'-\theta$, the direction of the
refracted wave vector is obtained through rotating the incident wave vector around $\mathbf{n}$ by
an angle $\Delta \theta$ within the frame $XYZ$. The wave number is changed by the refraction into
$k'= \frac{n_2}{n_1} k= n_2 k_0$. As a result, the refracted wave vector in the frame $XYZ$ is
given by
$$
\mathbf{k}'_{XYZ}=\frac{n_2}{n_1} D^{-1}_{\mathbf{n}}(\Delta \theta) \mathbf{k}_{XYZ},
$$
where
$$
D^{-1}_{\mathbf{n}}(\Delta \theta)
  =\left(\begin{array}{ccc}
            \cos \Delta\theta & -\frac{k_y}{k \sin\theta_0}(1-\cos \Delta\theta) & \sin \Delta \theta\\
           -\frac{k_y}{k \sin\theta_0}(1-\cos \Delta\theta) & 1 & \frac{k_y}{k\sin\theta_0} \sin \Delta\theta\\
           -\sin \Delta\theta & -\frac{k_y}{k\sin\theta_0} \sin \Delta\theta & \cos \Delta\theta
         \end{array}
   \right).
$$
Correspondingly, the mapping matrix of the refracted plane wave is obtained through the same
rotation and is given by
$$
m'_{XYZ}= D^{-1}_{\mathbf{n}}(\Delta \theta) m_{XYZ}.
$$

Now we look at the latter operation. Eq. (\ref{incidence MM in XYZ}) shows that the first and the
second column vectors of the mapping matrix are $p$- and $s$-polarized, respectively, in the
zeroth-order approximation. So the refraction converts the normalized Jones vector of the incident
beam into $T \tilde{\alpha}$, where $T=\left( \begin{array}{cc} t_p & 0 \\ 0 & t_s \end{array}
\right)$,
$$
t_p=\frac{2 \sin \theta'_0 \cos \theta_0}{\sin (\theta_0+ \theta'_0) \cos (\theta_0- \theta'_0)}
$$
is the Fresnel transmission coefficient of the $p$-polarization, and
$$
t_s=\frac{2 \sin \theta'_0 \cos \theta_0}{\sin (\theta_0+ \theta'_0)}
$$
is the Fresnel transmission coefficient of the $s$-polarization. Introducing the normalized Jones
vector,
\begin{equation}\label{refraction Jones-like vector}
\tilde{\alpha}'=\frac{1}{N}
                T \tilde{\alpha}
                \equiv \left( \begin{array}{c} \alpha'_p \\ \alpha'_s \end{array} \right),
\end{equation}
where $N=(|t_p \alpha_p|^2+|t_s \alpha_s|^2)^{1/2}$ is the normalization coefficient, the
amplitude scalar $f'$ in the electric vector (\ref{field vector in XYZ}) is given by
\begin{equation}\label{refraction amplitude scalar}
f'=N f.
\end{equation}
Eq. (\ref{refraction Jones-like vector}) means that the polarization ellipticity of the refracted
angular spectrum is different from that of the incident angular spectrum and turns out to be
\begin{equation}\label{refraction ellipticity}
\sigma'=\tilde{\alpha}'^{\dag} \hat{\sigma} \tilde{\alpha}'= \frac{t_p t_s}{N^2} \sigma,
\end{equation}
noticing that $t_p$ and $t_s$ are both real numbers.

\textit{3. Description of the refracted beam in frame $x'y'z'$.} The wave vector of the refracted
angular spectrum is transformed into
\begin{equation}\label{refracted wave vector in x'y'z'}
\mathbf{k}'
  \equiv \left( \begin{array}{c} k'_{x'} \\ k'_{y'} \\ k'_{z'} \end{array} \right)
  =D_y(\theta'_0) \mathbf{k}'_{XYZ}
  =\left(
      \begin{array}{c}
         \frac{\cos \theta_0}{\cos\theta'_0} k_x \\
          k_y                             \\
          k'
      \end{array}
   \right).
\end{equation}
The mapping matrix is transformed in the same way, $m'=D_y(\theta'_0) m'_{XYZ}$, and is expressed
in terms of $\mathbf{k}'$ as follows,
\begin{equation}\label{refraction MM in x'y'z'}
m'=
  -\left(
      \begin{array}{cc}
         1 & \frac{k'_{y'}}{k'} \frac{\cos\theta_0-\cos\theta'_0}{\sin\theta'_0} \\
        -\frac{k'_{y'}}{k'} \frac{\cos\theta_0-\cos\theta'_0}{\sin\theta'_0} & 1 \\
        -\frac{k'_{x'}}{k'} & -\frac{k'_{y'}}{k'}
      \end{array}
   \right).
\end{equation}
Finally, we arrive at the electric vector of the refracted angular spectrum in the frame $x'y'z'$,
\begin{equation}\label{refracted field vector in x'y'z'}
\mathbf{f}'=m' \tilde{\alpha}' f',
\end{equation}
where $m'$, $\tilde{\alpha}'$, and $f'$ are given by Eqs. (\ref{refraction MM in x'y'z'}),
(\ref{refraction Jones-like vector}), and (\ref{refraction amplitude scalar}), respectively.
Comparison of Eq. (\ref{refraction MM in x'y'z'}) with Eq. (\ref{linear MM}) shows that the
symmetry axis $\mathbf{I}'$ of the refracted beam lies in the plane $z'ox'$, that is to say, in
the incidence plane. The angle $\Theta'$ between $\mathbf{I}'$ and the propagation axis is
determined by
\begin{equation}
\cot \Theta'= \frac{\cos\theta_0-\cos\theta'_0}{\sin\theta'_0},
\end{equation}
and is no longer equal to $-\frac{\pi}{2}$.
\begin{figure}[ht]
\includegraphics{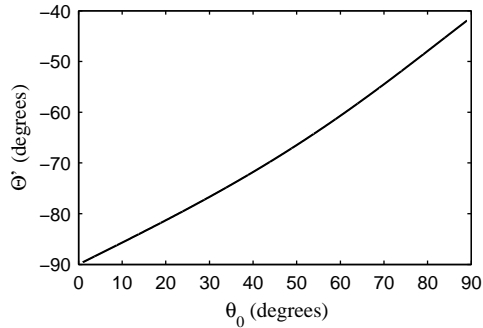}
\caption{Dependence of $\Theta'$ on the incidence angle $\theta_0$ for $n_1=1$ and $n_2=1.515$.}
\label{beam parameter}
\end{figure}
The dependence of $\Theta'$ on the incidence angle $\theta_0$ is shown in Fig. \ref{beam
parameter} for $n_1=1$ and $n_2=1.515$.

\begin{figure}[ht]
\includegraphics{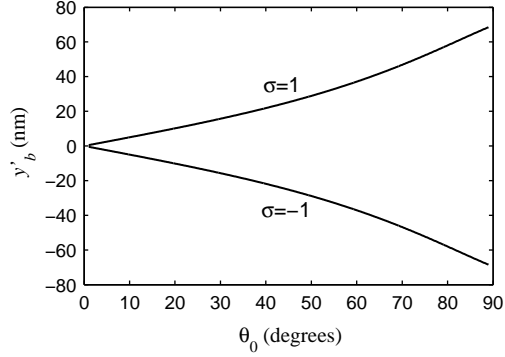}
\caption{Dependence of $y'_b$ on the incidence angle $\theta_0$ for experimental parameters:
$n_1=1$, $n_2=1.515$, and $\lambda_0=633$nm.} \label{IFD}
\end{figure}
According to Eq. (\ref{displacement}), the displacement of the refracted beam's barycenter from
the incidence plane is
\begin{equation} \label{IFeffect}
y'_b= -\frac{\sigma'}{k'} \cot \Theta'.
\end{equation}
This is nothing but the Imbert-Fedorov displacement. With $n_1=1$, $n_2=1.515$, and
$\lambda_0=633$nm, the dependence of $y'_b$ on the incidence angle $\theta_0$ for $\sigma= \pm 1$
is shown in Fig. \ref{IFD}, which is in quantitative agreement with the experimental data
\cite{Hosten-K}.

\textit{Indistinguishableness of different perpendicular $\mathbf{I}$}---When Hosten and Kwiat
performed their experiment, they did not realize the existence of $\mathbf{I}$. So it was not
ensured that the $\mathbf{I}$ of their incident beam lay in the incidence plane. In the following,
I will show that incident beams of different perpendicular $\mathbf{I}$ are indistinguishable in
the Imbert-Fedorov effect.

For an arbitrary perpendicular symmetry axis, $\mathbf{I}= -\mathbf{e}_x \cos \Phi- \mathbf{e}_y
\sin \Phi$, the mapping matrix in the linear approximation is given by
\begin{equation} \label{linear MM 2}
m= \left( \begin{array}{cc} -\cos \Phi &  \sin \Phi \\
                            -\sin \Phi & -\cos \Phi \\
                             \frac{k_x}{k} \cos \Phi+ \frac{k_y}{k} \sin \Phi & \frac{k_y}{k} \cos \Phi- \frac{k_x}{k} \sin \Phi
          \end{array}
   \right).
\end{equation}
Inserting a unit $2 \times 2$ matrix $I= Q^T Q$ into the right-hand side of Eq.
(\ref{factorization}), one has
\begin{equation} \label{insert}
\mathbf{f}=m Q^T Q \tilde{\alpha} f \equiv  m'' \tilde{\alpha}'' f,
\end{equation}
where
\begin{equation} \label{rotation matrix}
Q= \left( \begin{array}{cc} \cos \Phi & -\sin \Phi \\
                            \sin \Phi & \cos \Phi
          \end{array}
   \right)
\end{equation}
is a rotation matrix in the Jones-vector space, $m''= m Q^T$, and $\tilde{\alpha}''= Q \tilde{\alpha} \equiv \left( \begin{array}{c} \alpha''_p \\
\alpha''_s \end{array} \right)$. Eq. (\ref{insert}) means that the electric vector of the angular
spectrum can be expressed either in terms of $m$ together with Jones vector $\tilde{\alpha}$ or in
terms of $m''$ together with Jones vector $\tilde{\alpha}''$. With Eqs. (\ref{linear MM 2}) and
(\ref{rotation matrix}), one has
\begin{equation}
m''= -\left(
        \begin{array}{cc}
                1             &  0 \\
                0             &  1 \\
               -\frac{k_x}{k} & -\frac{k_y}{k}
        \end{array}
     \right),
\end{equation}
which is the same as the mapping matrix (\ref{incidence mm}). At the same time, rotation $Q$ does
not change the polarization ellipticity,
\begin{equation}
\sigma''= \tilde{\alpha}''^{\dag} \hat{\sigma} \tilde{\alpha}''= \tilde{\alpha}^{\dag} Q^T
\hat{\sigma} Q \tilde{\alpha}= \tilde{\alpha}^{\dag} \hat{\sigma} \tilde{\alpha}= \sigma.
\end{equation}
It is thus clear that when the perpendicular $\mathbf{I}$ of the incident beam is rotated around
its propagation axis, the linearly approximated mapping matrix and the polarization ellipticity
can be regarded as remaining unchanged, resulting in the same Imbert-Fedorov displacement
(\ref{IFeffect}).

This work was supported in part by the National Natural Science Foundation of China (60877055 and
60806041), the Science and Technology Commission of Shanghai Municipal (08JC14097 and 08QA14030),
the Shanghai Educational Development Foundation (2007CG52), and the Shanghai Leading Academic
Discipline Project (S30105).

\end{document}